\documentclass[superscriptaddress,twocolumn,prl]{revtex4-1}

\usepackage{graphicx}
\usepackage{amsmath}
\usepackage{amssymb}
\usepackage{amsthm}

\begin{document}

\title{Defect chemistry of Eu dopants in NaI scintillators studied by atomically resolved force microscopy}

\author{Manuel Ulreich}
\affiliation{Institute of Applied Physics, TU Wien, Wiedner Hauptstrasse 8-10/134, 1040 Vienna, Austria}

\author{Lynn A. Boatner}
\affiliation{Materials Science and Technology Division, Oak Ridge National Laboratory, Oak Ridge, Tennessee 37831, USA}

\author{Igor Sokolović}
\affiliation{Institute of Applied Physics, TU Wien, Wiedner Hauptstrasse 8-10/134, 1040 Vienna, Austria}

\author{Michele Reticcioli}
\affiliation{Institute of Applied Physics, TU Wien, Wiedner Hauptstrasse 8-10/134, 1040 Vienna, Austria}
\affiliation{University of Vienna, Faculty of Physics and Center for Computational Materials Science, Vienna, Austria}

\author{Berthold Stoeger}
\affiliation{X-ray Centre, TU Wien, Getreidemarkt 9, 1060 Vienna, Austria}

\author{Flora Poelzleitner}
\affiliation{Institute of Applied Physics, TU Wien, Wiedner Hauptstrasse 8-10/134, 1040 Vienna, Austria}

\author{Cesare Franchini}
\affiliation{University of Vienna, Faculty of Physics and Center for Computational Materials Science, Vienna, Austria}

\author{Michael Schmid}
\affiliation{Institute of Applied Physics, TU Wien, Wiedner Hauptstrasse 8-10/134, 1040 Vienna, Austria}

\author{Ulrike Diebold}
\affiliation{Institute of Applied Physics, TU Wien, Wiedner Hauptstrasse 8-10/134, 1040 Vienna, Austria}

\author{Martin Setvin}
\email{setvin@iap.tuwien.ac.at}
\affiliation{Institute of Applied Physics, TU Wien, Wiedner Hauptstrasse 8-10/134, 1040 Vienna, Austria}

\begin{abstract} 

Activator impurities and their distribution in the host lattice play a key role in scintillation phenomena. Here a combination of cross-sectional noncontact atomic force microscopy (nc-AFM), X-ray photoelectron spectroscopy (XPS), and density functional theory (DFT) was used to study the distribution of Eu$^{2+}$ dopants in a NaI scintillator activated by 3\% of EuI$_2$. A variety of Eu-based structures was identified in crystals subjected to different post-growth treatments. Transparent crystals with good scintillation properties contained mainly small precipitates with a cubic crystal structure and a size below 4~nm.  Upon annealing, Eu segregated towards the surface, resulting in the formation of an ordered hexagonal overlayer with a EuI$_2$ composition and a pronounced, unidirectional moire pattern. Crystals with poor optical transparency showed a significant degree of mosaicity and the presence of precipitates. All investigated crystals contained a very low concentration of Eu dopants present as isolated point defects; most of the europium was incorporated in larger structures. 

\end{abstract}

\maketitle

\section {Introduction}

Scintillating materials continue to attract increasing interest due to their use in medical instruments, radiation security, and scientific devices. One important class of scintillators is represented by halide materials activated by impurity dopants \cite{InorganicScintillators}. In particular, Tl-doped NaI \cite{Hofstadter1948} has maintained an important position in radiation detection systems for many decades, until it was largely superseded by Ce-doped LaBr$_3$ \cite{Loef2001}. In the last decade, the application of divalent Eu-doped halides such as SrI$_2$ have undergone rapid growth due to their excellent light yields; e.g., SrI$_2$:Eu$^{2+}$ can convert gamma rays or high-energy particles into blue scintillation light with impressive yields up to $\approx$100,000 photons/MeV \cite{Cherepy2009, Boatner2014SPIE, Boatner2015CaI2, Yokota2014}.

\begin{figure} [b]
    \begin{center}
        \includegraphics[width=1.0\columnwidth,clip=true]{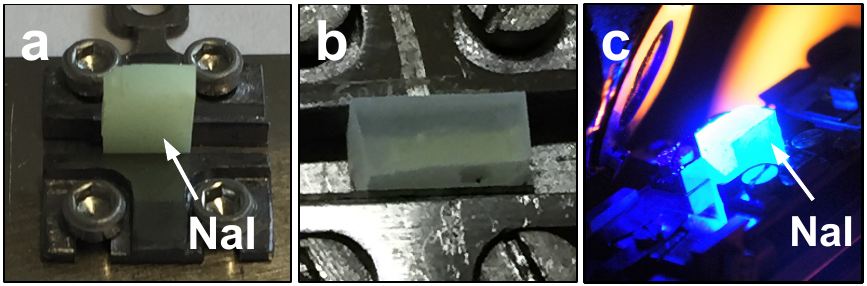}
    \end{center}
\caption{\textbf{NaI:3\%EuI$_2$ crystals.} a) As-grown crystal, mounted on a customized sample holder. b) Crystal subjected to post-growth annealing and quenching. c) Crystal scintillation during XPS measurements.}
\end{figure}

The scintillator performance is closely linked to the activator doping level: Higher doping levels can increase the scintillator light yield, thereby improving the energy resolution. However, at higher dopant concentrations, self-absorption phenomena, dopant segregation (precipitate formation), and other deleterious absorption phenomena can occur; this is especially prevalent when the dopant's charge state differs from the cations in the host crystal lattice. In particular, precipitate particles scatter scintillation light, decrease the optical transmission properties, and thereby reduce the radiation detection capabilities. Typical doping thresholds for precipitate formation can lie well below 1\% \cite{Guerrero1980}, and this process strongly depends on the crystal history, including treatments  such as annealing, quenching, or radiation absorption \cite{Guerrero1980}. Precipitates were initially studied by Suzuki in alkali halides doped by divalent impurities \cite{Suzuki1954, Suzuki1961}. A prototypical cubic structure called a Suzuki precipitate, based on a combination of substitutional cation dopants and cation vacancies, was identified by X-ray diffraction. Several studies have indicated the existence of other precipitate types \cite{Suzuki1955, Lopez1980, Salas1997, Savelev1974, Rubio1981, Boatner2018}, whose structures are not definitively established. 

Here, we focus on the defect chemistry of NaI crystals with a high Eu doping level of 3\%. The as-grown crystals are optically white and nontransparent for the scintillation light (see Fig. 1a), indicating the presence of precipitate particles. Post-growth annealing to 400$^\circ$C and rapid cooling in Ar or He vapor renders the material transparent (see Fig. 1b) and provides excellent scintillating properties. A similar post-growth treatment has previously been proven to be efficient for LiI:Eu \cite{Boatner2017, Boatner2018}. 

Currently there is relatively little experimental information regarding the atomic-scale structure of many scintillating centers.
The ability to characterize structural properties of scintillators is limited by their insulating character and low stability upon electron-beam irradiation. Here we use cross-sectional noncontact atomic force microscopy (nc-AFM) to probe the defect chemistry of as-grown and thermally treated NaI:Eu crystals. Recent improvements in the technique \cite{GiessiblReview, GiessiblPreamp} allow for investigations of even complex insulating materials, and for achieving atomic resolution of both point defects and precipitated structures. AFM has been employed previously for studying alkali halides doped by various divalent impurities \cite{Barth2008PRL, Barth2009PRL, Barth2009NJPhys, Meija2015}, and Suzuki precipitates with sizes of $\approx$100~nm were found. Here we show that the defect chemistry of the Eu$^{2+}$ dopants differs significantly from the previously reported results, and we relate the structure to the scintillator properties.

\begin{figure}
    \begin{center}
        \includegraphics[width=0.95\columnwidth,clip=true]{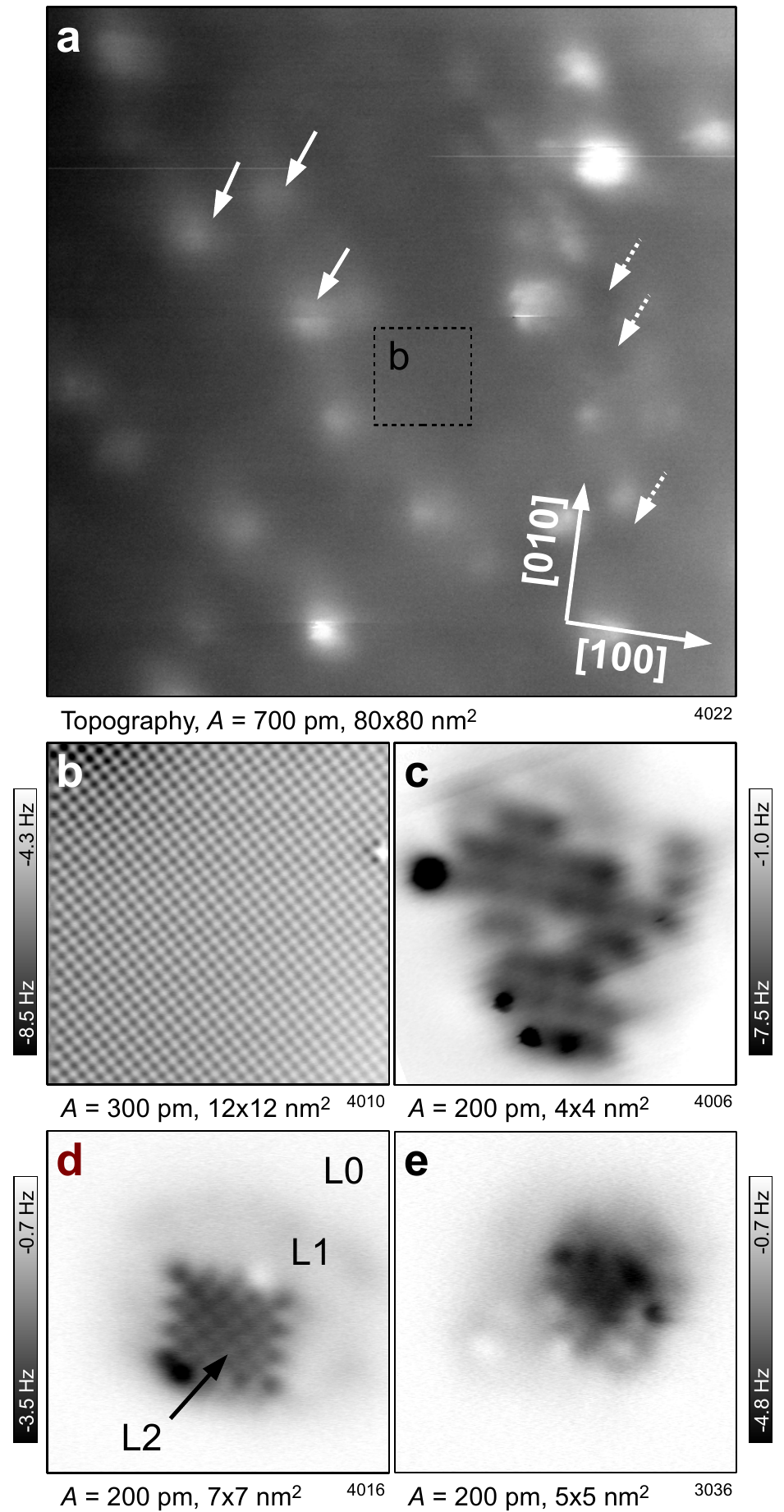}
    \end{center}
    \caption{
\textbf{Transparent NaI:3\%EuI$_2$ crystal quenched from 400$^\circ$C.}
a) Overview AFM image of a sample cleaved along the (001) surface at $T=-180^\circ$C. The dashed square marks a flat region imaged in b). The surface contains protrusions after the cleavage (marked by white arrows) and complementary depressions (dashed arrows). b) Atomically resolved AFM image showing the (1$\times$1) termination with a negligible concentration of point defects. c-e) Constant-height AFM images of the precipitates protruding from the cleavage plane. L0-L2 in d) denotes different atomic layers.  
}
\end{figure}

\section{Experimental details}

NaI:Eu samples were grown by the vertical Bridgman technique; details of the crystal growth and quenching are described in Ref. \cite{Boatner2017}. Nominal doping of 3 atomic percent EuI$_2$ was employed. After introducing the samples to ultrahigh vacuum (UHV) conditions in the AFM system, they were outgassed at 200-250$^\circ$C. Cleaving was performed along the (001) plane, using a tungsten-carbide blade. For all types of samples, the cleaving was done at several temperatures ranging from -180$^\circ$C to room temperature, in order to exclude the role of post-cleavage surface diffusion. 
The cleaved samples did not show any signatures of trapped point charges during AFM imaging, thus no post-cleavage annealing was performed, except where indicated in the text. Approximately 25 samples were analyzed by nc-AFM. 

AFM measurements were performed in an Omicron q-Plus LT head at $T=4.8$~K with a custom cryogenic preamplifier \cite{GiessiblPreamp}. Tuning-fork sensors ($f_0\approx31$ kHz, Q$\approx$20000) \cite{GiessiblPatent} with etched W tips were used, and the tips were cleaned by self-sputtering \cite{SetvinTips2012} and subsequently functionalized by touching the NaI (001) substrate. In all AFM images in the main text, atomic resolution stems predominantly from repulsive interaction above surface I atoms (bright spots). AFM imaging of ionic lattices is tip-dependent \cite{Barth2001, Shluger2003, Bechstein2009, Yurtsever2012, GrossNaCl}; we consistently present images measured with anion-terminated tips. Examples of images measured with a cation-terminated tip are shown in Fig. S1 (see supplemental material at   \cite{supplement}). A bright contrast in the AFM images can also correspond to a step down. Here a lower-lying atomic layer provides lower attractive forces in these constant-height AFM images.

X-ray photoelectron spectroscopy was performed with a non-monochromatized Al~K$\alpha$ line and a SPECS Phoibos~100 analyzer. The X-ray-induced photoconductivity is sufficiently high to reduce the sample charging to an acceptable level. Charging-induced peak shifts of up to 5~eV were detected, and the spectra presented here were shifted in a way that the Na 1s peak is aligned to 1071.4 eV.

X-ray diffraction measurements were performed on a Bruker Kappa APEX II diffractometer system using graphite-monochromatized MoKalpha radiation. Indexing and reciprocal space reconstruction were performed using the Apex3 software suite.

\section{Computational setup}

Calculations were performed by using the Vienna \emph{ab initio} simulation package (VASP)~\cite{Kresse1996a,Kresse1996}.
We adopted the generalized gradient approximation within the Perdew, Burke, and Ernzerhof (PBE) parametrization~\cite{Perdew1996}, revised for solids (PBEsol) \cite{PBEsol}.

We modeled the $Pnma$ and $P\bar{3}m1$ phases of EuI$_2$ with unit cells containing four Eu atoms and eight I atoms.
These structures were relaxed by using a plane-wave energy cutoff of 500~eV, and a $9\times 9\times 9$ sampling of the reciprocal space, which guarantees a precision of 1~meV per Eu atom.

Calculations of the substitutional Eu dopants in the cubic lattice were modeled using a 6$\times$6 unit cell, adopting the experimental lattice constant of 3.23 \AA. A slab thickness of 6 layers was used, with the bottom two layers fixed, and a vacuum gap of 13~\AA. The slab was relaxed using the Gamma point only, and the final energies were calculated using a 4$\times$4$\times$1 k-sampling. 

\section{Results and discussion}

\subsection{Cleaving transparent crystals}

\begin{figure}
    \begin{center}
        \includegraphics[width=1.0\columnwidth]{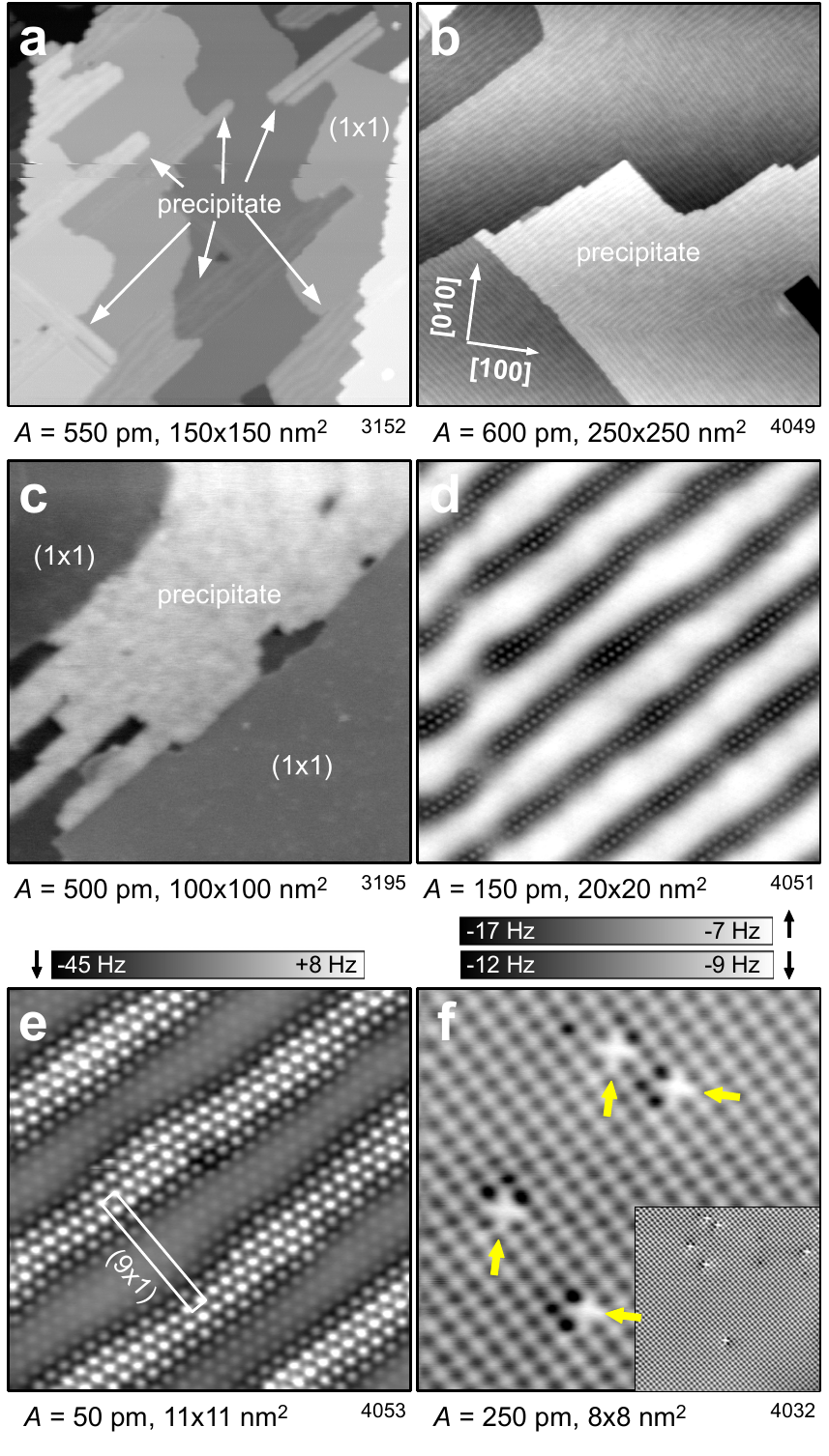}
    \end{center}
    \caption{
\textbf{Effect of annealing.} a-c) Overview AFM images of clear crystals annealed to a) 90$^\circ$C, b) 185$^\circ$C, and c) 250$^\circ$C, respectively. d) Constant-height image of the segregated phase, e) detail of the precipitate imaged at a tip-sample distance of $\approx$100~pm closer as compared to d). f) The (1$\times$1) phase (from a crystal annealed to 153$^\circ$C), showing point defects. Arrows highlight the different orientations of the defects. The inset shows a larger-scale area of the (1$\times$1) phase.  
}
\end{figure}

Figure 2 shows results of cross-sectional analysis of the transparent samples exposed to the post-growth treatment and quenching (see the photo in Fig. 1b). A representative overview AFM image of the cleaved sample is shown in Fig. 2a. The surface contains flat regions with the (1$\times$1) bulk-termination (see detail in Fig. 2b) and many small protrusions (marked by solid arrows). Complementary to the protrusions, depressions are also present (dashed arrows), yet are difficult to image in the noncontact AFM mode. The bulk-terminated regions exhibit a low defect density (below 0.3\%). Detailed images of the protrusions are shown in Figs. 2c-e. These features protrude typically 1-3 atomic layers above the surface plane and mostly retain the cubic crystal structure. The protruding parts are attributed to precipitates with an increased Eu concentration; the defect density in the flat regions is an order of magnitude below the nominal doping level, and the non-flat cleavage is clearly linked to the presence of Eu (cleavage of undoped samples generally provides atomically flat terraces). 


\subsection {Eu segregation upon annealing} 

Post-cleavage annealing was performed to gain qualitative information about bulk diffusion of the Eu dopants, and also to learn about possible precipitate phases. Previous works \cite{Barth2008PRL, Barth2009NJPhys} indicate that annealing of doped alkali halides results in dopant segregation towards the surface, and formation of surface precipitates with the same crystal structure as the bulk precipitates. 

The surface layer undergoes changes at surprisingly low temperatures, see Fig. 3. After annealing to 90$^{\circ}$C, elongated surface precipitates running along the [110] and [$\overline{1}$10] directions appear (Fig. 3a). In the temperature range from 150 to 200$^{\circ}$C, the surface becomes homogeneously covered by the precipitates (Fig. 3b). Surprisingly, the area of the precipitates decreases after further annealing above 200$^{\circ}$C; here the surface contains both, regions with the (1$\times$1) phase, and regions with the precipitates (Fig. 3c).

The detailed AFM images in Fig. 3d,e (more details are shown in Fig. S2) enable us to achieve an understanding of the nature of the precipitates. Even though the surface appears `striped' at larger tip-sample distances (Fig. 3d), a closer approach with the AFM tip shows that the hexagonal atomic arrangement is homogeneous across the whole surface and that the `stripe' pattern is a moire effect. We estimate the vertical corrugation of the moire as $\approx$120 pm and the moire periodicity as (9$\times$1) compared to the underlying (1$\times$1) bulk structure; the unit cell is marked in Fig. 3e. 

After annealing, the regions with the (1$\times$1) termination reproducibly show a specific type of point defects in a concentration of(0.7$\pm$0.2)\%, see Fig. 3f. These defects appear as bright spots with a darker region on one side (see the inset); four different orientations are possible. The origin and structure of these defects is discussed below.


\begin{figure}
    \begin{center}
        \includegraphics[width=1.0\columnwidth]{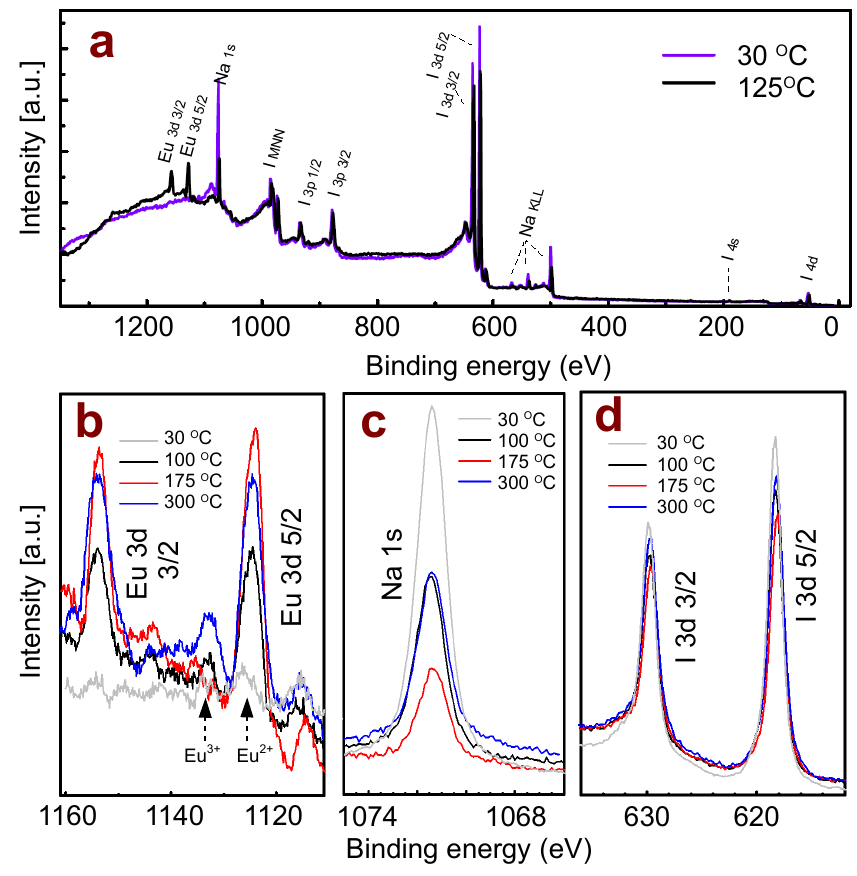}
    \end{center}
    \caption{
\textbf{XPS measurements.} a) Overview XPS spectra measured at room temperature and during annealing to 125$^\circ$C. b) Details of Eu~3d, c) Na~1s and d) I~3d peaks after annealing to various temperatures.
}
\end{figure}

The surface chemical composition at various stages of annealing was investigated by XPS, see Fig. 4. The overview spectra exclude any significant contamination -- only Na, I, and Eu are present. The spectra confirm Eu segregation towards the surface upon annealing. The signal of Na decreases with an increasing Eu content in the near-surface region, while the amount of Iodine does not undergo significant changes. Detailed XPS spectra (Fig. 4b-d) show that the Eu signal reaches its maximum at $\approx$175$^\circ$C and slightly decreases after further annealing, consistent with the AFM images in Fig. 3; there the area covered with the precipitate also first increases and then decreases.  

The strong decrease in the Na XPS signal indicates that the segregating phase is europium iodide. We attribute the remaining Na signal at the surface fully covered by the precipitate to a signal from subsurface layers. It seems plausible that a thin film of EuI$_2$ forms at temperatures below 200$^\circ$C. Annealing to higher temperatures could possibly change the character of the precipitate phase from 2D to 3D and this could explain why the (1$\times$1) phase reappears after annealing above 200$^\circ$C (Fig. 3c). All of the XPS data are consistent with the charge states of Na$^{1+}$, I$^{1-}$ and Eu$^{2+}$. The surface annealed to  300$^\circ$C shows a small peak at 1123~eV, which could be possibly interpreted as Eu$^{3+}$. 


\begin{figure}
    \begin{center}
        \includegraphics[width=1.0\columnwidth]{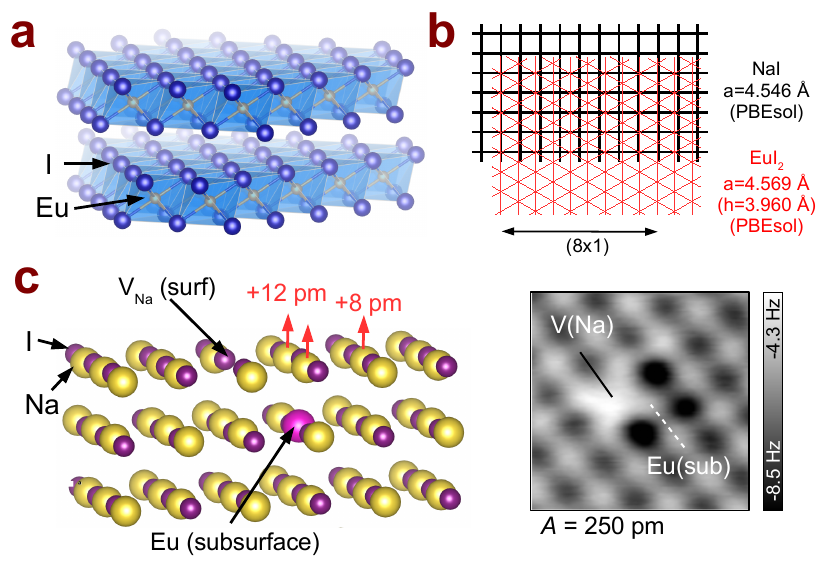}
    \end{center}
    \caption{
\textbf{Precipitate structure.} a) Proposed atomic structure of the EuI$_2$ precipitate phase. b) Schematic drawing of the moire pattern. The NaI (001) square lattice is drawn in black, the red grid denotes the hexagonal EuI$_2$ lattice with parameters calculated by DFT. The parameter $a$ denotes the length of the square, resp. triangle side. $h$ denotes the triangle height. c) Proposed structural model of the point defect shown in Figs. 2b and 3f. d) Detailed experimental image of the defect. The assumed position of the Na vacancy and subsurface Eu substitutional atom are marked. 
}
\end{figure}

\subsection {DFT calculations}

The precipitated phase observed in Fig. 3 is inconsistent with the cubic Suzuki structure, and also does not match any known bulk structure of EuI$_2$ \cite{Krings2009}. The moire pattern indicates that the precipitate is a 2D sheet, weakly bound to the substrate. We propose a structural model based on I-Eu-I trilayers: The related materials like CaI$_2$ \cite{Boatner2015CaI2, Gundiah2015} or CdI$_2$ \cite{Greenaway1965} have such layered bulk crystal structure (see Fig. 5a), where the trilayers are weakly coupled together. Similar layered structures are also common for many transition metal halides \cite{McGuire2017}. 

\begin{table}
\caption{\label{table1} Computed relative energies of various configurations of the Eu-V$_\mathrm{Na}$ defect}
	\centering
		\begin{tabular}{p{2.6in}p{0.6in}}
Configuration&	Energy\\
\hline
Eu(sub) + V$_\mathrm{Na}$(surf)	&	0\\ 
Eu(sub) + V$_\mathrm{Na}$(sub) &	+170 meV\\ 
Next-nearest neighbor Eu(sub) + V$_\mathrm{Na}$(sub) &	+206 meV\\ 
Eu(surf) + V$_\mathrm{Na}$(surf) &	+218 meV\\ 
Next-nearest neighbor Eu(surf) + V$_\mathrm{Na}$(surf) &	+298 meV\\ 
\hline
  \end{tabular}
\end{table}

We have calculated the stability of such a layered EuI$_2$ structure ($P3m1$) and compared it to the standard bulk structure of EuI$_2$ ($Pnma$). The calculations show that these structures are almost isoenergetic; the energy difference is 15~meV per EuI$_2$ unit (see Fig. S4 for details), and ferromagnetic ordering is preferred. The calculated lattice parameters match well with the observed moire pattern, see Fig. 5b. The layered EuI$_2$ structure has the iodine atoms arranged in a hexagonal lattice, with a I-I distance of 4.57~\AA. Overlaying this lattice (red) on the square NaI lattice (black) provides a moire pattern with a (8$\times$1) periodicity, while the experimentally observed pattern is (9$\times$1). The lattice mismatch along the perpendicular direction is 0.5\%, which allows for a commensurate arrangement and explains the stripe orientation along the $<$110$>$ directions. We note that the scheme in Fig. 5b uses lattice parameters calculated within the PBEsol approximation. When using differential computational setups or employing the experimmental NaI lattice constant (4.57~\AA), the moire periodicity ranges from 7.5 to 9.5 and the lattice mismatch ranges from 0.5 to 3.5\%.
  



The point defects observed at the surface (Fig. 2b and 3f) are consistent with the prototypical configuration expected for a single Eu$^{2+}$ dopant: It is widely accepted that Eu dopants occupy substitutional positions, replacing Na$^{+}$ cations. Such a defect must be coupled with a neighboring Na vacancy to maintain charge neutrality \cite{Savelev1974}. The defects in Fig. 3f fit this scheme well, and AFM images indicate that the Na vacancy prefers locations in the surface layer, while the Eu atom is located subsurface (see the model in Fig. 5c). 

We have calculated various near-surface configurations of such a Eu substitutional atom + Na vacancy, and the calculations clearly show that the Na vacancy is favored at the surface, while the Eu atom prefers a subsurface position (see Table 1, the corresponding coordinate files are attached as Supplementary Information). The energetically preferred configuration is intuitive, because the electrostatic charge of the Eu$^{2+}$ cation is better screened in a bulk-like environment, while the Na vacancy breaks fewer chemical bonds when present at the surface. The experimental AFM images (see the detail in Fig. 5d) fit well with the computed configuration: We attribute the central bright spot to the Na vacancy, which has a formal negative charge with respect to the unperturbed lattice, providing electrostatic repulsion with the iodine-terminated tip. The three faintly darker spots are attributed to vertical relaxation of the corresponding Na atoms; these are lifted by the underlying Eu$^{2+}$ ion by 12, 12, and 8~pm (see Fig. 5c). It is consistent with the slightly higher attractive force observed in AFM images. Furthermore, Kelvin Probe Force Microscopy (KPFM) spectra measured above these defects confirm the negative potential of the surface Na vacancy, forming a dipole with the Eu$^{2+}$ \cite{GrossNaCl, RubenNanoletters}, see Fig. S3. 

Importantly, our ability to observe point defects confirms that the transparent crystals investigated in Fig. 2 contain large Eu-free regions, $i.e.$, most of the Eu is incorporated in precipitates. 

\subsection {Cleaving opaque crystals}

\begin{figure}
    \begin{center}
        \includegraphics[width=1.0\columnwidth]{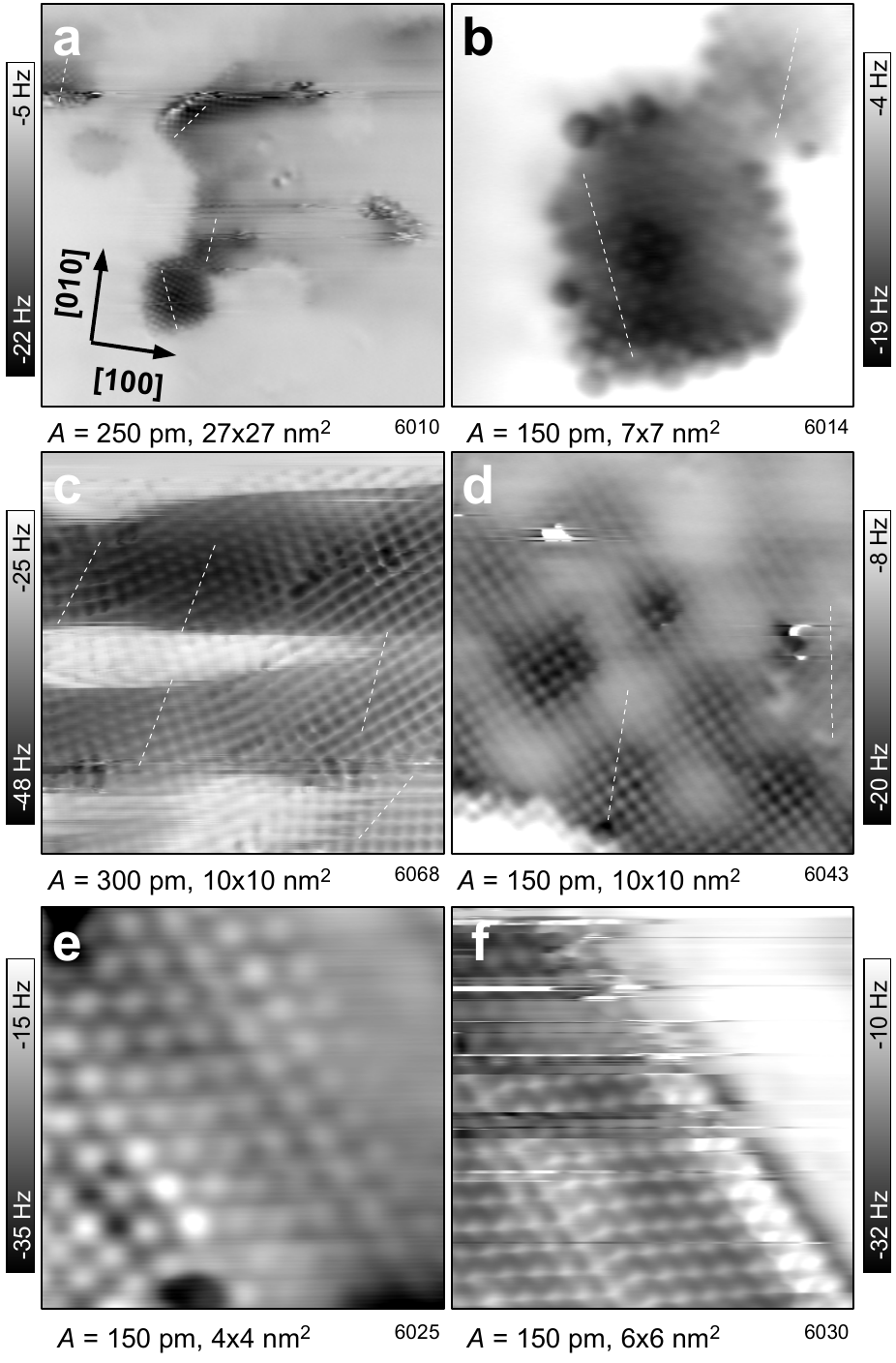}
    \end{center}
    \caption{
\textbf{AFM images measured on opaque NaI:3\%EuI$_2$ crystals cleaved at $-40^{\circ}$C.} a-c) Several regions showing a high level of mosaicity. Dashed lines mark the local orientation of the [010] direction. Arrows in a) denote the dominant orientation of the whole crystal, which is determined from the shape of cleaved crystals. d) Moire pattern with a square symmetry. e,f) Precipitates with the same structure as identified in Fig. 3 after annealing.
}
\end{figure}

An important question is - what is the reason of the opacity of the as-grown crystals shown in Fig. 1a. These crystals still exhibit a good cleavage plane, allowing for atomic-scale characterization by AFM. First, we note that the structure of the transparent crystals discussed above was very reproducible, $i.e.$, independent of the crystal history (e.g., shelf time, annealing) and the specific growth charge. In contrast, AFM imaging of the untreated (opaque) crystals showed a large variability. We noticed that even the standard outgasing cycle (30 min. at 200$^{\circ}$C) applied upon inserting the sample into UHV affects the crystals internal structure and defect distribution.

Our data indicate that a key feature in the opaque crystals is mosaicity - see Fig. 6. Figs. 6a-c show different regions of the crystal, where the cleavage plane shows many small crystallites rotated with respect to the main crystallographic axes of the crystal. The rotation is small (below 10$^{\circ}$) in most cases, and only few crystallites exhibit larger rotation. The rotation is apparent in both x-y plane, and in the local plane tilting. Interestingly, analysis of the AFM images indicates that all the crystallites have the same lattice constant (within the measurement error of $\approx3$\%), thus the driving force towards the mosaicity probably does not stem from lattice mismatch between the crystallites. 

Figure 6d shows another typical feature found in the opaque crystals: A moire pattern with square symmetry. We attribute this pattern to the overlay of two square lattices, which have the same lattice constant and are slightly rotated with respect to each other. This moire is therefore again attributed to the mosaicity and originates from a differently oriented crystallite located subsurface.  

Last, we have occasionally found areas with a hexagonal atom arrangement and a 1D moire (Fig. 6e,f). This structure seems identical to the layered EuI$_2$ precipitates discussed above. It indicates that this precipitate structure can indeed form in the bulk, but its occurrence on the as-cleaved surfaces was rather scarce. We note that the opaque crystals also contained regions where AFM imaging could not been performed; this indicates there are also other structures that do not possess a good cleavage plane.    

We have examined several of the Eu-doped NaI crystals by single-crystal X-ray diffraction, in order to check consistency with the cross-sectional AFM analysis. Figure 7a shows a (0$kl$) cut through the reciprocal space, measured on a transparent NaI crystal. Here a lower nominal Eu doping of 1.2\% was used to ensure that no precipitates are present in the crystal. The image shows the bulk diffraction spots expected for NaI. All other, weaker spots are artifacts originating from the monochromator: The radial elongation of the main diffraction spots is caused by a limited suppression of the Mo K$_{alpha2}$ line, and the half-order spots originate from X-rays with double energy of the Mo K$_{alpha}$. 

X-ray diffraction measured on the untreated crystals (Fig. 7b) clearly confirm the mosaicity observed in AFM: All the diffraction spots are elongated radially, which is a typical sign of mosaicity. These crystals also often show extra diffraction spots, indicating the presence of other precipitate types. However, a discussion of all possible structures appearing in the untreated crystals is beyond the scope of this paper. 

\begin{figure}
    \begin{center}
        \includegraphics[width=1.0\columnwidth]{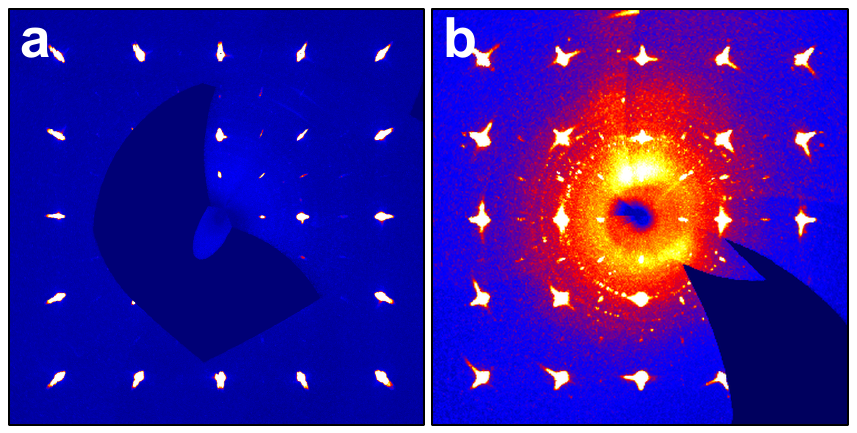}
    \end{center}
    \caption{
\textbf{XRD data.} a) $0kl$ cut through the reconstructed reciprocal space, measured on a low-doped (1.2\% Eu) NaI crystal rendered transparent by the post-growth treatment. b) The same data measured on an opaque, 3\%Eu-doped NaI cystal, without post-growth treatment.   
}
\end{figure}

\subsection{Relation between mosaicity and precipitation}

The origin of the mosaicity observed in the opaque crystals is somewhat puzzling. The crystallites have the same lattice constant as their parent matrix, as confirmed both by analysis of the AFM data and by the XRD measurements shown in Fig. 7b. We conclude that the crystallites have predominantly the NaI composition. The driving mechanism for the mosaicity therefore does not seem to originate from formation of bulk precipitates; it likely works in the opposite way. 

There is a clear correlation between the mosaicity and the Eu doping level. We hypothesize that the Eu dopants aggregate at grain boundaries of the crystallites. Such a mosaic structure could be a perfect starting point for formation of larger bulk precipitates. The perfect NaI lattice, characteristic for transparent crystals, probably creates a large barrier \cite{Science2018NucleationBarrier} for nucleation of precipitates, $i.e.$, overcoming certain critical size, where the energy gain would overcome the energy penalty for interfacing the NaI lattice. 

\section{Summary}

We have used cross-sectional AFM to analyze the defect chemistry of optically transparent and opaque NaI:3\%Eu crystals. In all cases, only a negligible concentration of point defects was detected; a majority of the Eu dopants was agglomerated in various precipitates. The transparent crystals contained small precipitates with sizes below 5~nm and a cubic crystal structure. Scintillation in these crystals, therefore, likely originates from precipitate structures, indicating that nanometer-sized precipitates are not an impediment to good scintillation properties.

A new type of precipitate structure was identified upon annealing the crystals, i.e., layered EuI$_2$ sheets with an I-Eu-I trilayer structure. While EuI$_2$ does not adopt such a layered structure in its bulk form, it can exist at surfaces, and possibly also inside a salt matrix. Several prior studies indicate the existence of similar precipitates in other systems \cite{Suzuki1955, Lopez1980, Salas1997, Savelev1974, Rubio1981, Boatner2018}, but their atomic structure has never been solved. This layered EuI$_2$ structure may be interesting for other applications beyond scintillation. The combination of the unique properties of Eu$^{2+}$, and the excellent optoelectronic properties of layered materials \cite{Jariwala2014} may provide interesting functionalities. Furthermore, layered metal iodides have recently attracted attention because of their magnetic properties \cite{McGuire2017, Botana2019, Klein2018, Huang2017, Jiang2018}. 

Comparisons of the as-grown crystals with crystals exposed to a post-growth treatment show that the type of precipitate structures can be effectively tuned. The exact type of the Eu precipitates formed in their parent matrix might be linked to subtle changes in the crystal chemical composition and trace impurities. 

\section{Ackonwledgment}

{This work was supported by the Austrian Science Fund (FWF) Project Wittgenstein Prize Z 250 and Solids4Fun W1243. Research at the Oak Ridge National Laboratory for one author (LAB) was supported by the US Department of Energy, Office of Science, Basic Energy Sciences, Materials Sciences and Engineering Division. The X-ray Center of the TU Wien is acknowledged for providing access to the single crystal diffractometer. The computational results have been achieved by using the Vienna Scientific Cluster (VSC).}
 



%


\end{document}